\documentclass[twocolumn,aps,prd,amssymb,eqsecnum,nofootinbib,showpacs,superscriptaddress]{revtex4-1} %

\usepackage{graphicx}
\usepackage{latexsym} 
\usepackage{color}
\usepackage{enumerate}		
\usepackage{hyperref}
\usepackage{epsfig}		
\usepackage{amsmath} 
\usepackage{amsfonts}
\usepackage{bbold}

\newcommand{\onefigure}[2]{\begin{figure}[htbp]
\begin{center}\leavevmode\epsfxsize=3.1in\epsfbox{#1.eps}\end{center}\caption{#2\label{#1}}
\end{figure}}
\newcommand{\twofigures}[3]{\begin{figure}[htdp]
\centering \leavevmode\epsfxsize=3.1in\epsfbox{#1.eps}
\leavevmode\epsfxsize=3.1in\epsfbox{#2.eps} 
\caption{{
#3}\label{#1}}
\end{figure}}

\def\be{\begin{equation}}			\def\ee{\end{equation}}
\def\bea{\begin{eqnarray}}		\def\eea{\end{eqnarray}}
\def\bw{\begin{widetext}}			\def\ew{\end{widetext}}

\def\C{{\cal C}}			
\def\lp{\left(}		\def\rp{\right)}		\def\lb{\left[}		\def\rb{\right]}
\def\non{\nonumber}

\begin{document}

\title{Coordinate families for the Schwarzschild geometry based on radial timelike geodesics}	

\author{Tehani K. Finch} 
\affiliation{NASA Goddard Space Flight Center\\
Greenbelt MD 20771}

\begin{abstract} 
We explore the connections between various coordinate systems associated with observers moving inwardly along radial geodesics in the Schwarzschild geometry.   Painlev\'e-Gullstrand (PG) time is adapted to freely falling observers dropped from rest from infinity; Lake-Martel-Poisson (LMP) time coordinates are adapted to observers who start at infinity with non-zero initial inward velocity; Gautreau-Hoffmann (GH) time coordinates are adapted to observers dropped from rest from a finite distance from the black hole horizon.  We construct from these an LMP family and a proper-time family of time coordinates, the intersection of which is PG time.	  We demonstrate that these coordinate families are distinct, but related, one-parameter generalizations of PG time, and show linkage to Lema\^itre coordinates as well.  
\end{abstract}

\maketitle			

\section{Introduction}
\label{sec:intro}
The Schwarzschild geometry is among the best known spacetimes of general relativity.  Not only is it an exact analytic solution of the Einstein equations, it has significant physical relevance as an excellent approximation to the spacetime outside the sun, and therefore as the starting point for many experimental tests of general relativity \cite{Har03}. The study of radial geodesics has long been a key tool for scrutinizing this spacetime, and coordinates adapted to null radial geodesics (ingoing and outgoing Eddington-Finkelstein (EF) coordinates) appear regularly in textbooks \cite{Har03}-\cite{Poi04}.  Coordinates adapted to \emph{timelike} radial geodesics have received much less attention, however, and indeed rarely appeared in the literature before the 2000s.  

Taylor and Wheeler in \cite{TW00} categorize radial timelike geodesics in terms of objects that are said to be in ``hail" frames, ``drip" frames, and the ``rain" frame.  Objects in a hail frame have been hurled inward toward the black hole from an effectively infinite distance;  they start at infinity with an initial inward velocity of magnitude $v_{\infty}$.   Objects in a drip frame are dropped from rest, from a finite initial radius $R_i$.  Between these two categories are objects in the rain frame, which can be thought of as having been dropped from rest, from infinitely far away.  Objects in the rain frame represent the $R_i\to \infty$ limit of the set of drip frames and the $v_{\infty}\to 0$ limit of the set of hail frames.  Apart from the treatments in \cite{MP01} and \cite{FK04}, coordinates adapted to these frames have seldom been discussed as a group.  This paper aims to provide a detailed clarification of the relationships between these sets of coordinates.

Meanwhile, the spatial volume contained within the event horizon of a black hole depends on the hypersurface of simultaneity used to compute it, and there are infinitely many possible values for this volume, each corresponding to a different time slicing.  Calculations of black hole volume can prove useful for recognizing the attributes of various coordinate systems, as shown by DiNunno and Matzner in \cite{DM10}.  In a similar spirit, the present study utilizes volume calculations to assist in the analysis of the coordinates affiliated with Schwarzschild radial timelike geodesics.  These include the Lake-Martel-Poisson (LMP) coordinates of the hail frames, the Painlev\'e-Gullstrand (PG) and Lema\^itre coordinates of the rain frame, and other coordinates, including those of Gautreau and Hoffmann (GH), of the drip frames (\cite{MP01}, \cite{PH21}-\cite{GH78}).  

In this paper, we lay the groundwork with a discussion of PG time $t_{PG}$ and its attributes in Section \ref{sec:PGtime}.  Then in section \ref{sec:LMP-GH-DF-time} we establish two separate extensions of PG time.  The first is LMP time $t_{LMP}$, but the second is shown not to be GH time $t_{GH}$, but rather an offshoot of it we dub $t_{DF}$ (drip frame proper time).  In section \ref{sec:completion} we present $t_{LMP*}$ (the drip-frame analog of LMP time) and $t_{HF}$ (hail frame proper time), which comprise the remaining pieces of a useful classification scheme.  In this scheme, $t_{LMP}$, $t_{PG}$, and $t_{LMP*}$ correspond to the hail frame, rain frame, and drip frame variants, respectively, of an ``LMP family" of time coordinates.  Similarly, $t_{HF}$, $t_{PG}$, and $t_{DF}$  correspond to hail frame, rain frame, and drip frame members of a ``proper-time family" of time coordinates.  The LMP and proper-time families represent two distinct generalizations of PG time. 
  
The coordinate systems discussed in Sections \ref{sec:PGtime} - \ref{sec:completion} make use of the familiar spherical coordinates $\{r, \theta, \phi\}$.  The term ``volume" will be used to refer to a spatial three-volume, meaning the proper volume of a region of a spacelike hypersurface.  In the examples considered below (with the exception of static time), surfaces of constant time are spacelike everywhere, penetrate the horizon, and extend all the way to the singularity.  

Henceforth units such that $G=c=1$ will be adopted.  Spacetime indices will be denoted by $\{a,b,c,...\}$.  A coordinate $\xi$ will be referred to as \emph{timelike} if surfaces of constant $\xi$ are spacelike hypersurfaces, and \emph{spacelike} if surfaces of constant $\xi$ are timelike hypersurfaces.  Brackets [] will be reserved for indicating the arguments of functions.
\section{The rain frame: Painlev\'e-Gullstrand coordinates} 
\label{sec:PGtime}
The familiar form of the Schwarzschild geometry with central mass $m$ is given with respect to static coordinates $\{t_s, r, \theta, \phi\}$:	
\be
ds^2 = -\lp 1- \frac{2m}{r} \rp dt_s^2  + \frac{ dr^2}{1-2m/r} + r^2 d\Omega^2. \label{staticmetric}	\ee	
\onefigure{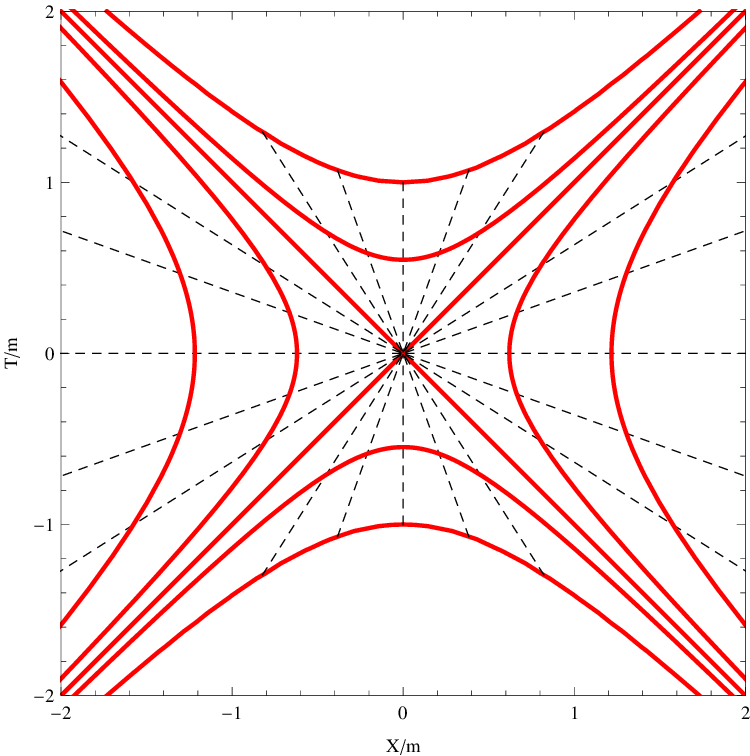}{Curves of constant $t_s$ (dashed) and constant $r$ (thick), for a Schwarzschild black hole.  The line $T=X$ corresponds to the event horizon $(r=2m, t_s=+\infty$) and the line $T=-X$ corresponds to the event horizon ($r=2m, t_s=-\infty$).  The uppermost hyperbolic curve represents the black hole/future singularity at $r=0$ and the bottom hyperbolic curve is the white hole/past singularity, also at $r=0$. }  
It is useful to visualize the behavior of time coordinates by means of a Kruskal diagram.  In such a diagram, null trajectories have slopes of $\pm 45^\circ$, and the horizontal and vertical axes refer to the Kruskal-Szekeres coordinates $X$ and $T$ respectively.  ``Our universe" corresponds to the portion of the spacetime above and to the right of the line $T=-X$.  The relation of Kruskal coordinates to static coordinates is discussed in, e.g., section 31.5 of  \cite{MTW73}.  Along with curves of constant $r$, curves of constant $t_s$ are displayed 
in Figure \ref{Fig-Statictime}. They do not penetrate the horizon, only approaching it asymptotically.  

The Schwarzschild geometry expressed in Painlev\'e-Gullstrand \cite{PH21,G22} coordinates \{$t_{PG}, r, \theta, \phi$\} 
is given by 
\be
ds^2 = -\left(1- \frac{2m}{r} \right) dt_{PG}^2+  2\sqrt{\frac{2m}{r}} dt_{PG}\, dr  
 +  dr^2 + r^2 d\Omega^2; \label{PGmetric}	
 \ee	
we can immediately note the convenient property that the spatial PG three-metric is flat.  
This time coordinate satisfies
\bea dt_{PG} &=& dt_s + \frac{\sqrt{2m/r}}{1-2m/r}dr, \label{dtPG} \\  
t_{PG} &=& t_s+2\sqrt{2m\,r} - 2m\ln \left| \frac{\sqrt{r/2m}+1}{\sqrt{r/2m}-1} \right| 	\label{PGtime}\\
&+& \C. 		\non
\eea
In (\ref{PGtime}) $\C$ refers to an arbitrary constant of integration.   The choice of  
$r=0$ as the reference point will be adopted for all coordinates studied in this paper with the exception of Gautreau-Hoffmann time $t_{GH}$.  The use of such a reference point implies that $\C=0$, and thus 
\bea
\lim_{r \to 0}\ &  t_{PG} & [t_s, r] = t_s,	\label{limit_PG}			
\eea
as seen in Figure \ref{Fig-Static-and-PGtime}.  (We note that Hamilton and Lisle in \cite{HL08} put forth a version of PG time that is defined with a reference point of $r=\infty$.  Their version would require $\C=-\infty$ in (\ref{PGtime}), which would make $t_{PG}$ take infinite negative values whenever both $r$ and $t_s$ are finite.  Hence, we do not make use of it.) 

The \emph{normal observers} of a foliation are those whose four-velocities $\vec{u}$ are orthogonal to the hypersurfaces of that foliation.  Here, the normal observers of the PG slicing have been dropped from infinity and are thus in the rain frame.  The observer's velocity is radial and given by 
\be \frac{dr}{d\tau} = -\sqrt{\frac{2m}{r}}.	\label{frefo}
\ee 
She also has a conserved energy-at-infinity per unit rest mass, $\tilde{E}$, associated with the timelike Killing vector of the Schwarzschild geometry.  For an observer in the rain frame $\tilde{E}=1$.  Intervals $\Delta \tau$ of proper time along an infalling $\tilde{E}=1$ geodesic correspond precisely to intervals of $t_{PG}$ \cite{Poi04,TW00}, so that $\Delta \tau = \Delta t_{PG}$.  
\onefigure{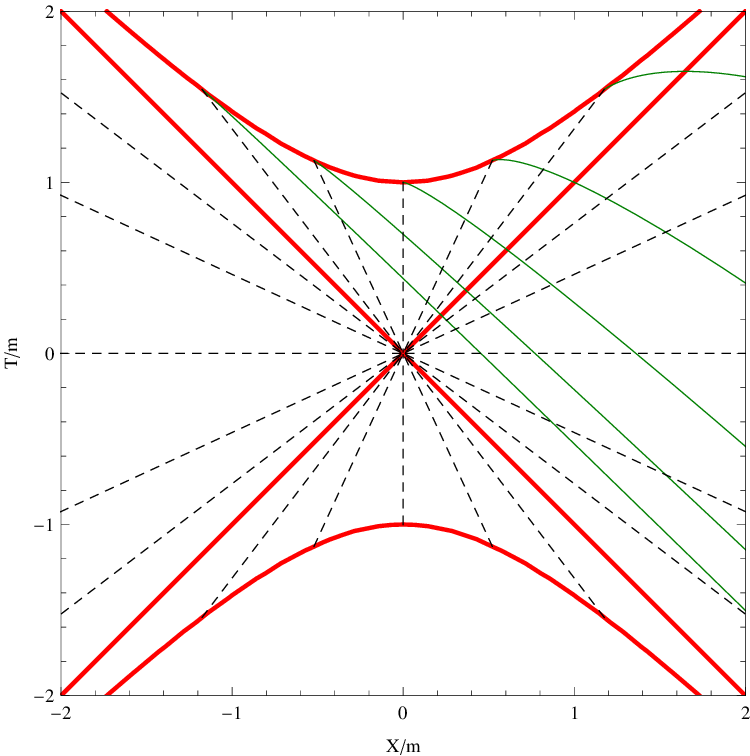}{Curves of constant $t_{PG}$ (thin) superposed onto curves of constant $t_s$ (dashed) and constant $r$ (thick).  From the left, the values shown are $t_{PG} = \{-4m,-2m,0,2m,4m \}$ and they intersect the corresponding curves $t_s = \{-4m,-2m,0,2m,4m \}$ at the (future) curve $r=0$.}  
The four-velocity of such an observer is therefore given by
\be
u^a := \frac{dx^a}{d\tau} = \lp 1, -\sqrt{ \frac{2m}{r} } , 0, 0  \rp \, .
\ee

We conclude this section by using PG coordinates to compute the black hole volume.  Since the determinant of the three-metric, $ {}^{(3)}g $, is $r^4\sin^2\theta$, the volume of a Schwarzschild black hole for the PG time slicing is simply 
\bea 
\mbox{Vol}_{PG} &=& \int^{2\pi}_{0}	  \int^{\pi}_{0}	\int^{r_{outer}}_{r_{inner}}   \sqrt{ {}^{(3)}g } \, dr \, d\theta \, d\phi \label{PGVol} 	 \\
&=&  4\pi\int^{2m}_0r^2 \,dr = \frac{4\pi}{3}(2m)^3=\frac{32\pi}{3}m^3, \non
\eea	
which happens to coincide with the familiar volume for a sphere of radius $2m$ in Euclidean three-space.  We will generalize (\ref{PGVol}) and provide hail-frame and drip-frame counterparts of many of these results in Sections \ref{sec:LMP-GH-DF-time} and \ref{sec:completion}.
\section{Coordinates based on general ingoing radial geodesics} 			
\label{sec:LMP-GH-DF-time}
Having given a flavor for the attributes of PG time, we now consider coordinates based on a broader class of trajectories.  The time coordinates discussed in this section are adapted to radial timelike geodesics of the Schwarzschild geometry for particles with general values of $\tilde{E}$.  
A convenient choice of parameter is $p=1/\tilde{E}^2$.  The $p<1$, $p=1$ and $p>1$ cases correspond to the hail frames, rain frame, and drip frames, respectively.  If $p<1$ one can associate $p$ with an initial velocity via $v_{\infty} = \sqrt{1-p}$; if $p>1$ one can associate $p$ with an initial radius via $R_i = 2Mp/(p-1)$.  The connections between these time coordinates (which were briefly treated in \cite{FK04} and in the endnotes of \cite{MP01}) will be used to group them into what we call the GH family, the LMP family, and the proper-time family. 
\subsection{	{\bf The hail frames: Lake-Martel-Poisson coordinates}	}	
\label{sec:LMPtime}
In \cite{MP01} Martel and Poisson analyzed an extension of the PG coordinate system, previously discovered by Lake \cite{L94}.  
They considered geodesic observers with initial inward velocity of magnitude $v_{\infty}$; these are the normal observers of the foliation and they have $\tilde{E}=1/\sqrt{1-v_\infty^2} = 1/\sqrt{p}$, where $0< p<1$.  The LMP time is not itself proper time of the normal observer with a given value of $p$; rather, intervals of LMP time are proportional to intervals of proper time $\tau^{(p)}$ via $\Delta t_{LMP}^{(p)}=\sqrt{p}\, \Delta \tau^{(p)}$.  The LMP time coordinates \{$ t_{LMP}^{(p)}$\} provide a straightforward extension of PG time to the $p<1$ cases.

Explicitly, in LMP coordinates \{$ t_{LMP}^{(p)},r,\theta,\phi $\}, the four-velocities of these observers are given by
\be
u^a = \lp \sqrt{p}, 
- \sqrt{\frac{1}{p} - \lp 1- \frac{2m}{r} \rp } \, , 0, 0  \rp \, .	\label{u_LMP}
\ee 
The LMP time coordinate satisfies\footnote{The first term in parentheses of (\ref{LMPtime}) corrects a slight error in the corresponding equation of \cite{MP01} in which the square root was omitted.}	
\bw 	
\bea 
dt_{LMP}^{(p)} &=& dt_s + \frac{\sqrt{1-p(1-2m/r)}}{1-2m/r}\, dr, \label{dtLMP}\\	
t_{LMP}^{(p)} &=& t_s + 2m \biggl( \frac{ r\sqrt{1-p(1-2m/r)} }{2m} + \ln \biggl| \frac{1 - \sqrt{1-p(1-2m/r)}}{1 + \sqrt{1-p(1-2m/r)}} \biggr| 
 \nonumber \\ & & \mbox{    }  
- \frac{1-p/2}{\sqrt{1-p}} \ln \biggl| \frac{\sqrt{1-p(1-2m/r)}	
-\sqrt{1-p}}{\sqrt{1-p(1-2m/r)}+\sqrt{1-p}} \biggr| \, \biggr), \label{LMPtime}
\eea	
\ew				 
and also (dropping the superscript)
\bea 
\lim_{r\to 0} & t_{LMP} & [t_s,r]  = t_s\, .		\label{limit_LMP} 	
\eea
The Schwarzschild line interval in the LMP coordinates is given by
\bea
ds^2 &=& -(1-2m/r)\, dt_{LMP}^2 + p\, dr^2	\label{LMPmetric} 	 \\
&+& 2\sqrt{1-p(1-2m/r)}\, dt_{LMP}\,dr  + r^2\, d\Omega^2 .	\non		
\eea
\onefigure{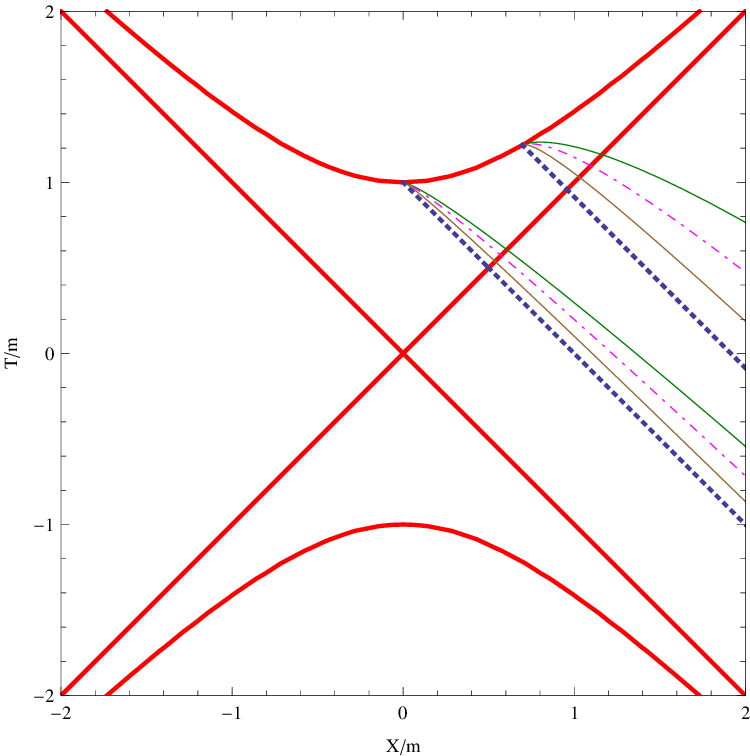}{Curves of constant $t_{LMP}$, shown along with thick dashed null curves of constant EF coordinate $v$, which is the $p\to0$ limit of LMP time, for comparison.  The curves within a bundle all meet at $r=0$.  The lower bundle corresponds to $t_{LMP}= 0$, the upper bundle to $t_{LMP}= 2.6m$.  Within each bundle, from the bottom, the values of $p$ are $\{0,1/3, 2/3, 1\}$, where the $p= 1$ case is PG time.} 	
In contrast to (\ref{PGmetric}), for $p\neq 1$ the spatial three-metric in (\ref{LMPmetric}) (and those for the rest of the coordinates in this section) is curved, as can be verified through a calculation of the three-dimensional Riemann tensor.  It is useful to keep in mind that each value of $p$ represents a different coordinate, and is thus associated with an entire foliation of the spacetime.  Figure \ref{Fig-LMPtime} shows that ``bundles" of curves that correspond to a given value of $t_{LMP}$, but with different values of $p$, merge at $r=0$.  This is due to the fact that (\ref{limit_LMP})
holds regardless of $p$.

Turning our attention to the boundaries of the above range of $p$: in the limit $p\rightarrow 1$, we obtain PG time \cite{MP01}.  The limit $p\rightarrow 0$ results in the ingoing Eddington-Finkelstein (EF) coordinate $v$:  
\bea	
dv &=& dt_s + \frac{dr}{1-2m/r}, \\
v &=& t_s + r + 2m\ln \biggl| \frac{r}{2m} - 1 \biggr|, \\
ds^2 &=& -(1-2m/r)\, dv^2 + 2dv\,dr +  r^2\, d\Omega^2 \label{in-EF-metric}.
\eea
Slices of constant $v$ are null; they follow the paths of ingoing radial light rays.  This makes $v$ itself a null coordinate.  Thus, strictly speaking, $v$ is not a time and is excluded from the LMP family introduced in Section \ref{sec:lmp-family}.  

If $p<1$, use of $t_{LMP}$ specifies a different surface of simultaneity from that of $t_{PG}$, and the result for the volume inside a Schwarzschild black hole generalizes:
\be
\mbox{Vol}_{LMP} 
=\int \sqrt{ {}^{(3)}g } ~ d^3 x = 4\pi\int^{2m}_0 \sqrt{p}\,r^2\, dr = \frac{32\pi\sqrt{p}}{3}m^3.
\label{LMPVol} \ee
This volume becomes arbitrarily small as $p\rightarrow 0$, $i.e.$ as $v_{\infty}\rightarrow 1$. 
\subsection{	{\bf The drip frames: Gautreau-Hoffmann coordinates	} 	}
\label{sec:DFtime}
Coordinates inspired by the perspective of observers dropped from rest a finite distance from the black hole, i.e. in a drip frame, were introduced by Gautreau and Hoffmann in \cite{GH78}.\footnote{The GH time coordinate should not be confused with that introduced by Novikov in \cite{Nov63}, even though both are derived from $\tau^{(R_i)}$ of (\ref{tau-Ri}).  The GH time coordinate is related to proper time readings of observers dropped from the \emph{same} location at \emph{different} times; vice versa for the Novikov time coordinate.  Slices of constant Novikov time are in fact very different from those of constant GH time. }  
In this subsection we consider their coordinate system, which turns out not to relate to the PG coordinate system in the desired manner.  Thus we will have to alter their time coordinate to arrive at a coordinate family that is a suitable outgrowth of PG time.  

Freely falling observers dropped from rest at $r=R_i$ have a conserved energy-at-infinity per unit rest mass $\tilde{E}=\sqrt{1-2m/R_i}\,$.
The proper time that elapses along their trajectories can be given in terms of $r$ 
(the following is a slight modification of that given in Chapter 26 of \cite{Blau14} 
and is equivalent to that in Chapter 31 of \cite{MTW73}):  
\be	
\tau^{(R_i)}=\frac{R_i^{3/2}}{(2m)^{1/2}}\lp\arccos \lb\sqrt{\frac{r}{R_i} }\,\rb+\sqrt{\frac{r}{R_i}-\lp \frac{r} {R_i}\rp^2 } \,\rp.
\label{tau-Ri}
\ee
In this scheme, $\tau^{(R_i)} =0$ corresponds to the instant at which the observer is dropped from $r=R_i$.
 
The quantity $\tau^{(R_i)}$, as given in (\ref{tau-Ri}), will not lead us to a suitable time coordinate; were we to attempt to use it, slices of constant $\tau^{(R_i)}$ would be slices of constant $r$ and hence timelike hypersurfaces.  Instead, the most useful form of $\tau^{(R_i)}$ is
\be \tau^{(R_i)} 
= \kappa\, t_s + h[r] \, ,	
\ee
where $\kappa$ is a constant and h[r] is often called the ``height function."  This is the type of transformation that keeps the metric stationary and the spherical symmetry manifest, as discussed in \cite{FK04}.  To obtain $\tau^{(R_i)}$ in this form, Gautreau and Hoffmann, rather than utilizing (\ref{tau-Ri}) directly, start with the geodesic equation.  They arrive at the four-velocity for an observer in a drip frame in static coordinates  \{$ t_s,r,\theta,\phi $\}:	
\be
u^a = \lp \frac{	\sqrt{1-2m/R_i}	}{1-2m/r},
- \sqrt{	\frac{2m}{r}- \frac{2m}{R_i}  } \, , 0, 0  \rp \, .	\label{u_stat}
\ee 
Equation (\ref{u_stat}) in turn gives
\be \frac{dt_s}{dr} = \frac{dt_s/d\tau^{(R_i)}}{dr/d\tau^{(R_i)}} = -\frac{\sqrt{1-2m/R_i}	}{(1-2m/r)\sqrt{2m/r-2m/R_i}}, \label{dts-dr} 
\ee
and after some algebra, the combination of (\ref{u_stat}) and (\ref{dts-dr}) yields 
\be
\frac{d\tau^{(R_i)}}{dr} = \sqrt{1-2m/R_i}\,\frac{dt_s}{dr} + \frac{\sqrt{2m/r-2m/R_i}}{1-2m/r}\, . \label{dtau-dts-Ri}	\ee
They then choose the initial location to be $\{t_s=0, r=R_i\}$.  This specifies a fiducial trajectory, and (\ref{dtau-dts-Ri}) now implies that the points \emph{along this trajectory} satisfy
\be
\tau^{(R_i)} = \sqrt{1-2m/R_i}\,\int_{0}^{t_s} d\tilde{t} + \int_{R_i}^r	\frac{\sqrt{ 2m/\tilde{r}-2m/R_i } }{1-2m/\tilde{r} }\, d\tilde{r} .	\label{tau-ts-Ri}
\ee 
\onefigure{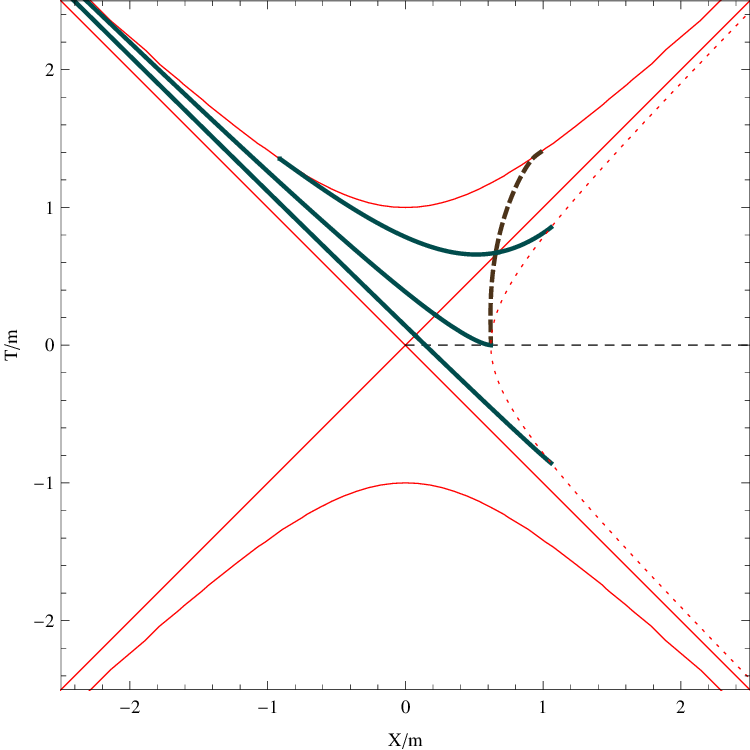}{Curves of constant ${t_{GH}}^{(R_i)}$ for a particular value of $R_i$.  Here, $R_i = 2.25m$.  The constant-time curves (thick solid) are for ${t_{GH}}^{(R_i)} = \{-1.5m, 0, 1.5m \}$.  Also pictured is the trajectory (thick dashed) of an observer dropped from $r =R_i$ at $t_s=0$.    The hypersurface $t_s=0$ (the portion of it outside the horizon, in our universe) is the thin dashed horizontal line, and the dotted hyperbola indicates $r=2.25m$.} 
The next step is to introduce a \emph{general} coordinate ${t_{GH}}^{(R_i)}$ that is allowed to take positive and negative values, and is valid for points both on and off the fiducial trajectory.  This coordinate is defined such that the transformations between $t_s$ and ${t_{GH}}^{(R_i)}$ take the same form as those for $\tau^{(R_i)}$ in (\ref{dtau-dts-Ri}) and (\ref{tau-ts-Ri}):
\bw  
\bea 
d{t_{GH}}^{(R_i)} &=& \sqrt{1-2m/R_i}\, dt_s + \frac{\sqrt{ 2m/r-2m/R_i } }{1-2m/r}\, dr, \label{dtgh-Ri}	
\\
{t_{GH}}^{(R_i)} &=& \sqrt{1-2m/R_i}\,t_s	 \label{tgh-Ri} 	\\
&+&	\sqrt{ \frac{2m}{R_i}} \biggl(\sqrt{r(R_i-r)}+(R_i-4m)\lp \arctan\lb \sqrt  { \frac{r}{R_i-r} }~ \rb -\frac{\pi}{2} \rp	\non \\ 	
& & -\sqrt{2m(R_i-2m)}\ln \biggl| \frac{ \sqrt{r(R_i-2m)}+\sqrt{2m(R_i-r)} }	{ \sqrt{r(R_i-2m)}-\sqrt{2m  (R_i-r)}  }  \biggr|\ \biggr).  	\non
\eea  
\ew
Here $R_i$ is treated as a constant in (\ref{dtgh-Ri}) and (\ref{tgh-Ri}), and both integrals in (\ref{tau-ts-Ri}) have been evaluated explicitly.  The Gautreau-Hoffmann time ${t_{GH}}^{(R_i)}$ is only defined for $r\le R_i$, and furthermore it is assumed that $R_i>2m$.  Some curves of constant ${t_{GH}}^{(R_i)}$ are plotted along with the corresponding fiducial trajectory in Figure \ref{Fig-GH-trajectory} for the case of $R_i = 2.25m$.  The line interval has the form (dropping the superscript)
\bea 
ds^2  
&=&-\frac{ 1 }{1-2m/R_i } \lp 1- 2m/r  \rp dt_{GH}^2   \label{GHmetric}\\
&+&  \frac{ 2\sqrt{ 2m/r - 2m/R_i } }{1-2m/R_i}dt_{GH}\, dr+\frac{dr^2}{1-2m/R_i} + r^2 d\Omega^2. \non 
\eea   

Unlike the other coordinates that have been discussed, ${t_{GH}}^{(R_i)}$ is defined with respect to a reference point of $r=R_i$ (as opposed to $r=0$) in the sense that 
\be \lim_{r \to R_i} {t_{GH}}^{(R_i)}[t_s = 0, r] = 0.	\label{limit_tau-Ri}
\ee
Equation (\ref{limit_tau-Ri}) is consistent with the fact that the hypersurface ${t_{GH}}^{(R_i)}= 0$ starts on the line $t_s=0$ for any $R_i>2m$; a few examples are shown graphically in Figure \ref{Fig-GHtime}.
\onefigure{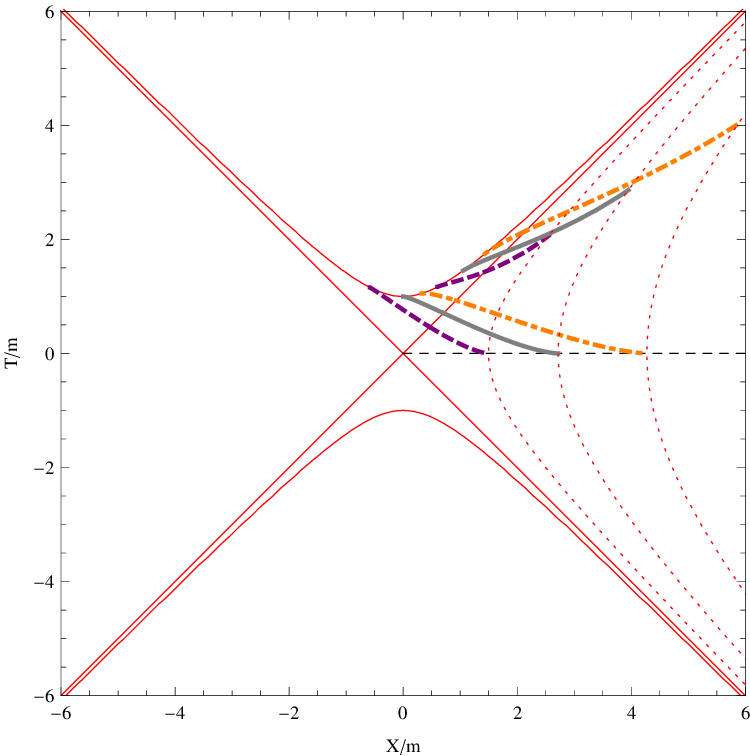}{Curves of constant ${t_{GH}}^{(R_i)}$ for various values of $R_i$.  The lower bundle corresponds to ${t_{GH}}^{(R_i)}= 0$, the upper to ${t_{GH}}^{(R_i)}=2.6m$.  Within each bundle, the values of $p$ are 3 (thick dashed), 2 (thick solid), and 5/3 (thick dot-dashed); these curves correspond to $R_i = 3m, 4m$, and $5m$ respectively. (The $p=1$ case is not well defined for this coordinate.)  The hypersurface $t_s=0$ (the portion of it outside the horizon, in our universe) is the thin dashed horizontal line on which the curves of the lower bundle begin, at $r=R_i$.  Plots of $r=3m, 4m$, and $5m$ (the dotted hyperbolae) are included for clarity.}  
However, this implies that when we consider the $r\to 0$ limit of ${t_{GH}}^{(R_i)}$, a slight complication arises:
\be \lim_{r\to 0} {t_{GH}}^{(R_i)}[t_s = 0, r] =\pi (4m-R_i)\sqrt{\frac{m}{2R_i} } \neq 0,	\label{limit_tau-0}
\ee
in contrast to the behavior of $t_{PG}$ in (\ref{limit_PG}) and $t_{LMP}$ in (\ref{limit_LMP}).  Equation (\ref{limit_tau-0}) implies that the $R_i\to \infty$ limit of ${t_{GH}}^{(R_i)}$ does not have a well-defined correspondence with PG time ( this would be no different had we chosen any other finite value of $\C$ in (\ref{PGtime}) ); this limit differs from PG time by an infinite constant.\footnote{Another way of stating this is that the difficulty with the $R_i\to \infty$ limit of ${t_{GH}}^{(R_i)}$ is due to the infinite transit time from $r=\infty$ to finite values of $r$.  Reference \cite{GH78} points this out, although without explicitly mentioning PG time.  Now, this particular difficulty could be avoided if one were to start with the Hamilton-Lisle version of PG time from \cite{HL08}, but we choose not to do so for reasons given in Section \ref{sec:PGtime}.}  Therefore, $t_{GH}$ is \emph{not} the $p>1$ analog of $t_{PG}$ that we seek.

To construct a coordinate that possesses the desired correspondence with PG time, and can easily be compared to LMP time, we keep (\ref{limit_tau-0}) in mind and introduce 
\be {t_{DF}}^{(R_i)}: = {t_{GH}}^{(R_i)} - \pi (4m-R_i)\sqrt{\frac{m}{2R_i} }. \label{tDF-Ri}
\ee
As was the case with ${t_{GH}}^{(R_i)}$, surfaces of constant ${t_{DF}}^{(R_i)}$ cannot be extended outward past $r=R_i$.  Since these geodesics are those of observers in drip frames, we will refer to $ {t_{DF}}^{(R_i)}$ as ``drip-frame proper time" (or DF time). 

The advantage of using $R_i$ is that the allowed range of the $r$ coordinate is clear.  However, to facilitate comparison between coordinates, and the formulation of coordinate families, we will present our expressions in terms of $p$.   We recall that if $p>1$, 
\be R_i = 2m\, p/(p-1) \leftrightarrow p = \frac{R_i}{R_i-2m}.	
\label{Ri-to-p}
\ee 
The range $\infty>R_i>2m$, corresponds to $1<p<\infty$.   Combining (\ref{tDF-Ri}) with (\ref{Ri-to-p}) tells us that
\be 
{t_{DF}}^{(p)} = {t_{GH}}^{(p)} -  \frac{\pi m (p-2)}{\sqrt{p(p-1)}}.		\label{tDF-subtraction}
\ee
Naturally, since ${t_{DF}}^{(p)}$ and $t_{GH}^{(p)}$ differ only by an additive constant, intervals of ${t_{DF}}^{(p)}$ still have a correspondence with the proper time of an observer dropped from $r=R_i$, in that ${\Delta t_{DF}}^{(p)} = \Delta\tau^{(p)}$ between two given events.

The \emph{GH family} of time coordinates we define to include the set $\{{t_{GH}}^{(p)} \}$; the set of time coordinates $\{{t_{DF}}^{(p)}\}$ will later be incorporated into the \emph{proper-time family}.\footnote{Although in this paper the designations ``drip-frame time" and ``proper-time family" are associated only with ${t_{DF}}^{(p)}$, it should be emphasized that physically, \emph{both} $\Delta t_{GH}^{(p)}$ and  $\Delta {t_{DF}}^{(p)}$ correspond to proper time intervals of an observer in a drip frame.} Notable from (\ref{tDF-Ri}) is the fact that ${t_{DF}}^{(p)}$ reduces to ${t_{GH}}^{(p)} $ for the case $R_i = 4m$, which corresponds to $p=2$.  A comparison of Figure \ref{Fig-GHtime} to Figure \ref{Fig-DFtime} helps illustrate that the $\{{t_{DF}}^{(p)}\}$ \emph{as a set} has a nontrivial difference from the set $\{{t_{GH}}^{(p)}\}$.  One example of this is that these two figures have opposite ``ordering" of the $p=5/3$, $p=2$ and $p=3$ curves within the bundles.  

In DF coordinates, the four-velocity of radial geodesic observers can be written as
\bea
u^a &=& \lp 1, -\sqrt{ \frac{2m}{r}-\frac{2m}{R_i}} , 0, 0  \rp	\non	\\
&=&	\lp 1, -\sqrt{ \frac{1}{p}-\lp 1- \frac{2m}{r} \rp } , 0, 0  \rp .		\label{u_DF}
\eea
\onefigure{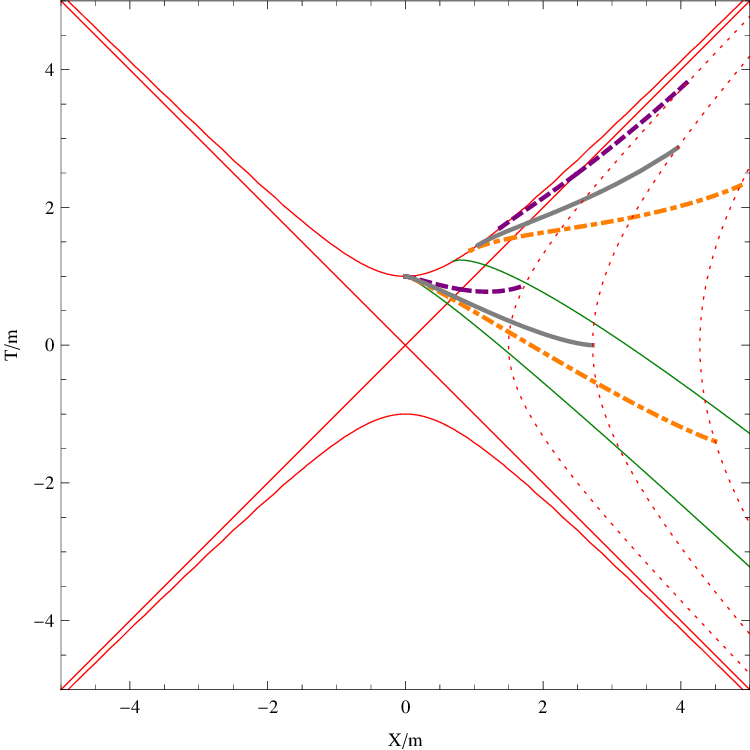}{Curves of constant $t_{DF}$.  Unlike Figure \ref{Fig-LMPtime}, here the curves of the upper bundle do not meet at $r=0$.  The lower bundle corresponds to $t_{DF}=0$, the upper to $t_{DF}=2.6m$.  Within each bundle, from the bottom, the values of $p$ are $\{1, 5/3, 2, 3\}$.  Plots of $r=3m, 4m$, and $5m$ (the dotted hyperbolae) are included for clarity.  The $p=1$ case is PG time. }%
The coordinate transformations for ${t_{DF}}^{(p)}$ are (dropping the superscript)
\bea	
dt_{DF} &=& \frac{dt_s}{\sqrt{p} } +\frac{1}{\sqrt{p}}\frac{ \sqrt{1-p(1-2m/r)} }{1-2m/r} dr,   \label{dtDF-alt} \\
t_{DF} &=& \frac{1}{\sqrt{p}}\Bigg( t_s +  r\sqrt{1-p(1-2m/r)} 	\label{tDF-alt} \\ 
&-& \frac{2m(p-2)}{\sqrt{p-1} } \arctan\lb	\sqrt{  \frac{p-1}{1-p(1-2m/r)} 	}\ \rb \non \\
&-& 2m\ln \Bigg| \frac{1+\sqrt{1-p(1-2m/r)}}{1-\sqrt{1-p(1-2m/r)}}\Bigg| \Bigg).		\non
\eea
The expression for the metric\footnote{Although they do not give it in explicit form, the authors of \cite{BGJ} show that it is possible to arrive at the metric of (\ref{DFmetric-p}), as well as the PG metric (\ref{PGmetric}), through analysis of the action for geodesics in Schwarzschild spacetime.} 
becomes 
\bea	ds^2 &=& -p \lp 1- 2m/r  \rp dt_{DF}^2  + p \,dr^2  \label{DFmetric-p}\\
&+&  2\sqrt{p} \sqrt{1-p(1-2m/r)}\, dt_{DF}\, dr + r^2 d\Omega^2.	\non	
\eea
Like $t_{PG}$ and $t_{LMP}$, $t_{DF}$ is defined relative to $r=0$, and the $p\to 1$ limit of $t_{DF}$ is in fact PG time.  

Comparing (\ref{dtDF-alt}) with (\ref{dtLMP}) shows that $dt_{DF}$ and $dt_{LMP}$ have a similar form, but the factor of $1/\sqrt{p}$ that distinguishes them has nontrivial consequences: equation (\ref{tDF-alt}) implies that
\be
\lim_{r\to 0} {t_{DF}}^{(p)}[t_s,r]  = \frac{t_s}{ \sqrt{p} }\ ,		\label{limit_tDF} 
\ee
in contrast to (\ref{limit_LMP}).  The fact that the right-hand side of (\ref{limit_tDF}) depends on $p$ implies that curves of constant $t_{DF}$ within the same bundle will not in general converge to the same $\{X,T\}$ point on the $r=0$ surface, contrary to what occurs in bundles of $t_{LMP}$ (this can be seen by comparing the upper bundles of Figure \ref{Fig-LMPtime} and Figure \ref{Fig-DFtime}).  
The spatial volume of the black hole for a slice of constant $t_{DF}$ has the same form as that in (\ref{LMPVol}):
\be
\mbox{Vol}_{DF}  =\int \sqrt{ {}^{(3)}g }\, d^3 x = 4\pi\int^{2m}_0 \sqrt{p}\,r^2\, dr 
=\frac{32\pi\sqrt{p} }{3}m^3.	\label{DFVol} 
\ee
This volume becomes arbitrarily large as $p$ approaches $\infty$, $i.e.$ as $R_i\rightarrow 2m$.

For completeness we note that since, for any $p$, $d{t_{GH}}^{(p)} = d{t_{DF}}^{(p)}$, the metric written in terms of $p$ for GH coordinates has the same form as (\ref{DFmetric-p}), and the black hole volume for a slice of constant ${t_{GH}}$ is also the same:
\bea	ds^2 &=& -p \lp 1- 2m/r  \rp d{t_{GH}}^2  \label{taumetric-p}  \\
 +  2&\sqrt{p}& \sqrt{1-p(1-2m/r)}\, d{t_{GH}}\, dr + p \,dr^2 + r^2 d\Omega^2,	 \non \\
 \mbox{Vol}_{GH}  &=& 4\pi\int^{2m}_0 \sqrt{p}\,r^2\, dr = 32\pi\sqrt{p}\,m^3/3.	\label{tauvolume-p}	
 \eea
\section{Completing the LMP and proper-time families} 
\label{sec:completion}
\subsection{	{\bf A $p>1$ analog of $t_{LMP}$}	} 
\label{sec:lmp-family}
The next step in constructing our coordinate families is obtaining a drip-frame analog of the set of LMP times.  As with ${t_{DF}}^{(p)}$, these coordinates cannot be extended past $r= 2m\,p/(p-1)$, and they can be expressed in terms of $R_i$ but not $v_{\infty}$.  The $p>1$ version of LMP time we denote as $t_{LMP*}^{(p)}\,$.  The infinitesimal coordinate transformation between $dt_s$ and $dt_{LMP*}$ is the same as in (\ref{dtLMP}), but, since $p>1$, the finite coordinate transformation differs from (\ref{LMPtime}):  
\bea 
dt_{LMP*}^{(p)}&=&	
dt_s + \frac{\sqrt{1-p(1-2m/r)}}{1-2m/r}\, dr,	\label{dtGF}\\		
t_{LMP*}^{(p)} &=& 
  t_s +  r\sqrt{1-p(1-2m/r)}  \label{GFtime}\\
  &-& \frac{2m(p-2)}{\sqrt{p-1} } \arctan\lb 	\sqrt{  \frac{p-1}{1-p(1-2m/r)} 	} \  \rb \non	\\
&-&	2m\ln \Bigg| \frac{1+\sqrt{1-p(1-2m/r)}}{1-\sqrt{1-p(1-2m/r)}}\Bigg| . 		\non
\eea
\onefigure{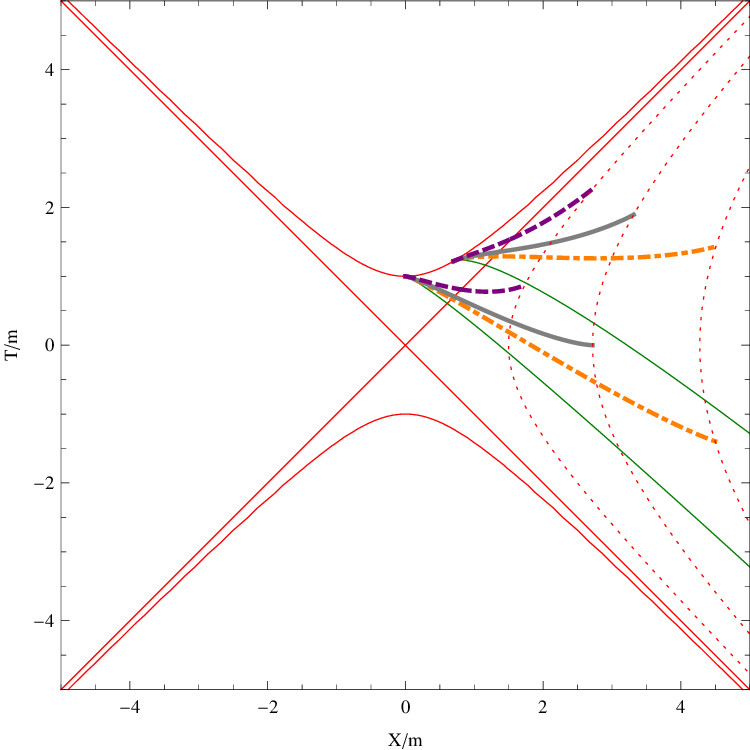}
{Curves of constant $t_{LMP*}$.  The lower bundle corresponds to $t_{LMP*}=0$, the upper bundle to $t_{LMP*}=2.6m$.  Within each bundle, from the bottom, the values of $p$ are $\{1, 5/3, 2, 3 \}$, where the $p=1$ case is PG time.  Plots of $r=3m, 4m$, and $5m$ (the dotted hyperbolae) are included for clarity.} 
In fact, comparison with (\ref{tDF-alt}) shows that formally 
\be t_{LMP*}^{(p)} = \sqrt{p}\, {t_{DF}}^{(p)}.	\label{LMP*relation}	
\ee

Thus, for any $p$, a slice of constant $t_{LMP*}^{(p)}$ is a slice of constant ${t_{DF}}^{(p)}$ and vice versa.  Equation (\ref{LMP*relation}) indicates that for a \emph{given} value of $p$, the coordinate $t_{LMP*}^{(p)}$ differs from the corresponding ${t_{DF}}^{(p)}$ only by an overall factor.  Yet \emph{as a collective}, the set $\{t_{LMP*}^{(p)} \}$ differs from $\{ {t_{DF}}^{(p)} \}$ nontrivially.  For example, curves of constant $t_{LMP*}^{(p)}$ within the same bundle converge at $r=0$ (just as occurs with ${t_{LMP}}^{(p)}$) while those of constant ${t_{DF}}^{(p)}$ in general do not.  This contrast becomes evident in a comparison of the upper bundle of Figure \ref{Fig-DFtime} with that of Figure \ref{Fig-LMPstar}. 

The metric with respect to $t_{LMP*}$ is given by 
\bea
ds^2 &=& -(1-2m/r)\, dt_{LMP*}^2 	\\
&+& 2\sqrt{1-p(1-2m/r)}\, dt_{LMP*}\,dr + p\, dr^2 + r^2\, d\Omega^2, \non
\eea
the same form as (\ref{LMPmetric}).  The normal observers have a four-velocity identical to that in (\ref{u_LMP}).  The evaluation of the black hole volume along slices of constant $t_{LMP*}$ gives   
\be
\mbox{Vol}_{LMP*} = 4\pi\int^{2m}_0 \sqrt{p}\,r^2 \, dr
=\frac{32\pi\sqrt{p} }{3}m^3.	\label{GFVol}
\ee 
Curves of constant $t_{LMP*}$ are shown by themselves in Figure \ref{Fig-LMPstar} and together with curves of constant $t_{LMP}$ in the top panel of Figure \ref{Fig-LMP_family}.  Establishing $t_{LMP*}$ as the $p>1$ analog of $t_{LMP}$ allows us to see $t_{LMP}^{(p)}$, $t_{PG}$, and $t_{LMP*}^{(p)}$ as encompassing, respectively, the $0<p<1$, $p=1$, and $1<p<\infty$ constituents of a one-parameter family of coordinates, which we call the \emph{LMP family}.  Appendix \ref{four-velocity} discusses how the time coordinates in this family are related to the four-velocities of normal observers via
\be
u_{a} = -\frac{1}{\sqrt{p}}\partial_{a} t \, .	\label{4-vel-01}
\ee  
\subsection{	{\bf A $p<1$ analog of $t_{DF}$}	}
\label{sec:gh-family}
Finally, we provide a hail-frame analog of $t_{DF}$.  Recall that the distinguishing feature of DF time is that ${\Delta t_{DF}}^{(p)} = \Delta\tau^{(p)}$ for an observer with $p>1$.  The desired analog \,${t_{HF}}^{(p)} $\, (with ``HF" indicating ``hail frame") would have ${\Delta t_{HF} }^{(p)}={\Delta \tau}^{(p)}$ for an observer moving along an inward geodesic with $p<1$, and would generalize a key property of the PG coordinate system.  
\onefigure{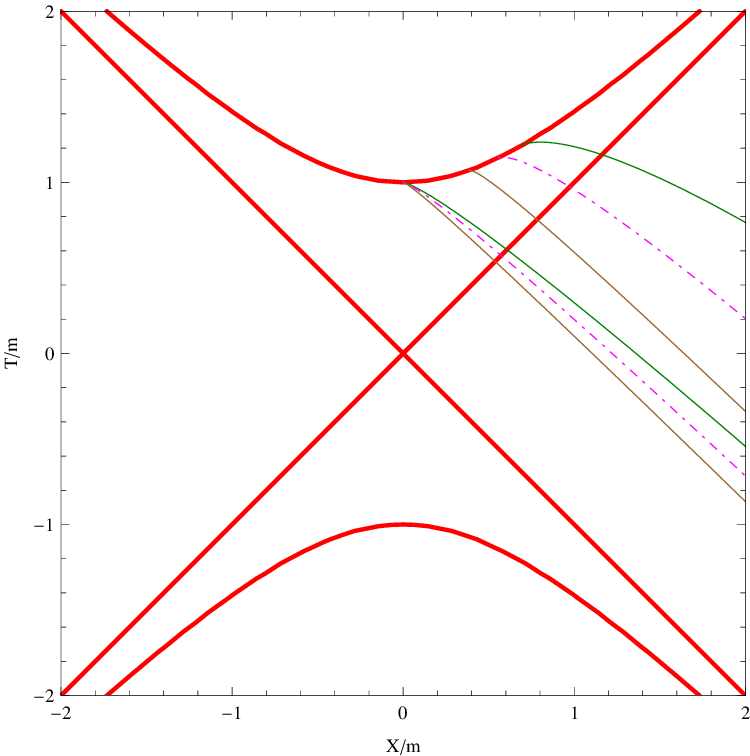}	
{Curves of constant $t_{HF}\,$. The lower bundle corresponds to $t_{HF}= 0$, the upper bundle to $t_{HF}=2.6m$.  Within each bundle, from the bottom, the values of $p$ are $\{1/3, 2/3, 1\}$, where the $p=1$ case is again PG time.}  
Such a coordinate ${t_{HF}}^{(p)}$ is given by 
\be {t_{HF}}^{(p)}  = \frac{  t_{LMP}^{(p)}  }{\sqrt{p}}, \label{HFrelation}
\ee
where $t_{LMP}^{(p)} $ is given in (\ref{LMPtime}), and satisfies
\bea	
d{t_{HF}}^{(p)}  &=&  
\frac{dt_s}{\sqrt{p} } +\frac{1}{\sqrt{p}}\frac{ \sqrt{1-p(1-2m/r)} }{1-2m/r} dr,   \label{dTee}
\eea	 
\twofigures{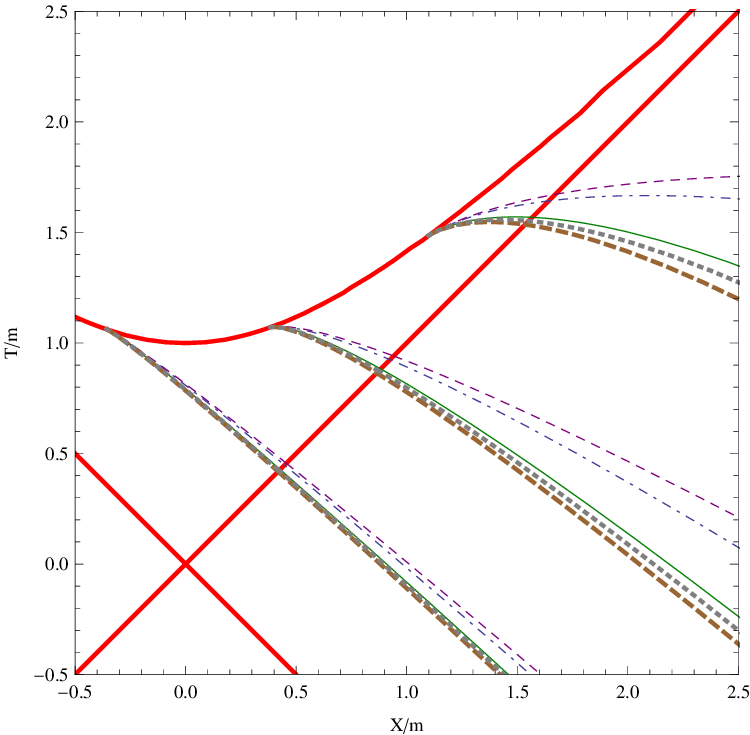}{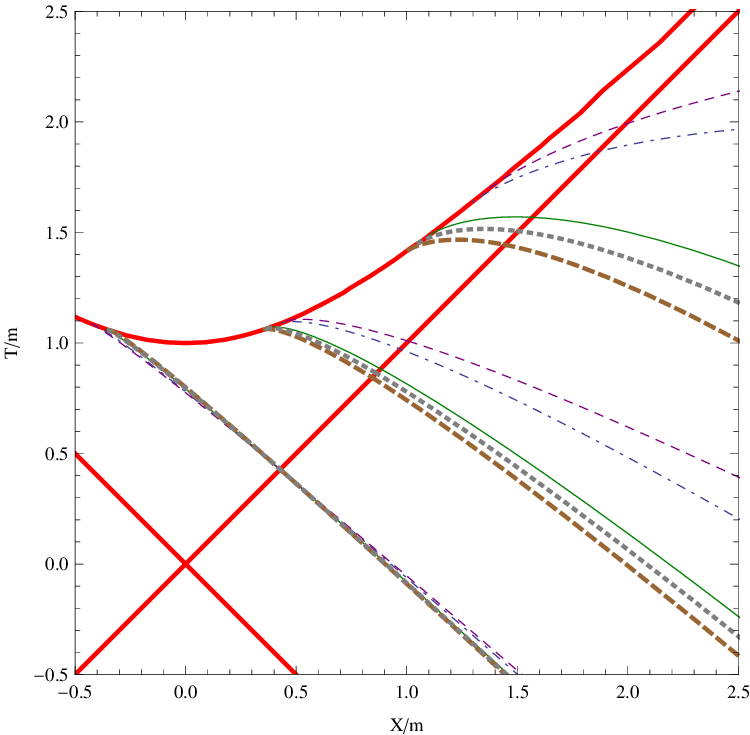}{Top panel: Plots from the LMP family, containing curves of constant $t_{LMP}$ along with curves of constant $t_{LMP*}$ and constant $t_{PG}$.  The lower bundle corresponds to $\{t_{LMP},t_{PG},t_{LMP*}\}=-1.4m$; the middle and upper bundles to  $1.5m$ and $3.8m$ respectively.  Within each bundle, from the bottom, the values of $p$ are $\{7/8, 15/16,1,9/7, 7/5 \}$, and within any given bundle the curves do not cross.  Bottom panel: Plots from the proper-time family.  The lower bundle corresponds to $\{t_{HF},t_{PG},t_{DF}\}=-1.4m$; the middle and upper bundles to $1.5m$ and $3.8m$ respectively.  Each bundle contains a range of $p$ values given by, from the bottom, $\{7/8, 15/16,1,9/7, 7/5 \}$.  For the proper-time family, when a bundle corresponds to a negative value of time, its curves \emph{do} cross, as can bee seen in the lower bundle.} 
the same form as (\ref{dtDF-alt}).  As noted in Section \ref{sec:LMPtime}, the $p\to 0$ limit of $t_{HF}$ is not well defined.  This is because the $p \to 0$ limit is associated with null trajectories, and proper time is not defined on a null trajectory.  Curves of constant $t_{HF}$ are plotted by themselves in Figure \ref{Fig-HFtime} and together with curves of constant $t_{DF}$ in the bottom panel of Figure \ref{Fig-LMP_family}.  

The resulting metric is given by
\bea
ds^2 &=& -p(1-2m/r)\, dt_{HF}^2 \label{Teemetric} \\ 
&+& 2\sqrt{p}\sqrt{1-p(1-2m/r)}\, dt_{HF} \, dr + p\, dr^2 + r^2\, d\Omega^2, \non
\eea
and the four-velocity of the normal observers is identical to that in (\ref{u_DF}).  The volume computation reproduces the familiar result  
\bea  
\mbox{Vol}_{HF} &=& 4\pi\int^{2m}_0 \sqrt{p}\,r^2\, dr = \frac{32\pi\sqrt{p}}{3}m^3.
\eea

The time intervals $\Delta {t_{HF}}^{(p)}$, $\Delta t_{PG}$, and  $\Delta {t_{DF}}^{(p)}$ all coincide with the proper time intervals of their respective normal observers; thus ${t_{HF}}^{(p)}$, $ t_{PG}$, and  $ {t_{DF}}^{(p)}$ correspond to the $0<p<1$, $p=1$, and $1<p<\infty$ members of a \emph{proper-time family}.   They are all defined with respect to a reference point of $r=0$, and they all satisfy 
\be
u_{a} = -\partial_{a} t,				\label{4-vel-02}
\ee  
on which we elaborate in Appendix \ref{four-velocity}.  Thus the proper-time family constitutes perhaps the most direct generalization of PG coordinates.    

Despite their simplicity, the fact that the relations (\ref{LMP*relation}) and (\ref{HFrelation}) depend on $p$ has significant consequences.  These include the contrast between the top and bottom panels of Figure \ref{Fig-LMP_family}; for the LMP family the curves within a bundle do not cross each other and converge to a point at $r=0$.  In the case of the proper-time family, the curves within a bundle do cross if the bundle corresponds to a negative value of the time coordinates, and generally do not meet at $r=0$ (the exception being the $t=0$ bundle).  There is also the fact that the LMP family has a well-defined $p\to 0$ limit, but the proper-time family does not.  Furthermore, only for the LMP family do the $g_{tt}$ components of the metrics enjoy the same form, namely $-(1-2m/r)$, that is familiar from static coordinates.  It is thus seen that although the LMP and proper-time families are related, there are still disparities between them.  
%
\section{Discussion}
\label{Discussion}
It is now well understood that, due to general covariance, the predictions of general relativity for behavior of a physical system are independent of the coordinates used to describe them.  Nonetheless, the choice of coordinates can be very significant, because this choice often has much influence on both the manageability of the calculations and on the amount of physical insight gained from them.  Examples in which PG coordinates have proven useful include the ``river model" of Schwarzschild spacetime in which space itself flows inward toward the horizon, through a flat background \cite{HL08}; the description of Hawking radiation as a tunneling process \cite{PW00}; and analytic models of gravitational collapse \cite{ABCL05}.  
Meanwhile, GH coordinates have been exploited for comparing massive particle trajectories in black hole and wormhole spacetimes \cite{Pop10}; in addition they have inspired the discovery of GH-type coordinates for de Sitter spacetime \cite{Gau83}, which have been utilized in the description of a Schwarzschild mass \cite{Gau84b} (and more recently a Reissner-Nordstr\"om charged mass \cite{Pos14}), embedded in a cosmological background.  	

With those developments as a backdrop, the present exposition has examined time coordinates adapted to general ingoing timelike radial geodesics in the Schwarzschild geometry.  To obtain proper-time analogs of PG time for the drip frames and hail frames, we have introduced the coordinates ${t_{DF}}^{(p)}$ and $ {t_{HF}}^{(p)}$, respectively; to our knowledge, analysis of these coordinates has not occurred elsewhere.  We have also provided a drip-frame analog of LMP time in $t_{LMP*}^{(p)}$.  The fact that the result $\frac{32\pi}{3}\sqrt{p}\,m^3$ for the Schwarzschild black hole volume is valid for all of the coordinate systems studied exhibits the close relation between them.  

We have chosen to group these coordinates into a GH family $\{{t_{GH}}^{(p)} \}$, an LMP family $\{t_{LMP}^{(p)}, t_{PG}, t_{LMP*}^{(p)} \}$ and a proper-time family $\{{t_{HF}}^{(p)}, t_{PG}, {t_{DF}}^{(p)} \}$.  The proper-time family intersects the GH family when $p=2$, and intersects the LMP family for the case of $p=1$ (PG time).  In particular, the LMP and proper time families represent a successful familial classification of two distinct one-parameter generalizations of Painlev\'e-Gullstrand coordinates. 
\section*{Acknowledgments} 
The author gratefully acknowledges fruitful correspondence with Brandon DiNunno and Richard Matzner; commentary from Tristan Hubsch and Bernard Kelly; discussions with James Lindesay that introduced him to Painlev\'e-Gullstrand coordinates; support from the Howard University Department of Physics and Astronomy, where this work was begun; and support from a NASA Postdoctoral Fellowship through the Oak Ridge Associated Universities.  
%
\appendix
\section{Lema\^itre coordinates: A time-dependent metric adapted to the rain frame} 
\label{Lemaitre}
It is instructive to explore the matters of Sections \ref{sec:PGtime} for a coordinate system which gives the Schwarzschild metric an explicit time dependence.  This scheme begins by noting that the motion of an infalling $\tilde{E}=1$ particle can be determined by solving (\ref{frefo}), with the result being that the trajectory $r[\tau]$ takes the form
\be
r[\tau]=\lp\frac{3}{2}\sqrt{2m}(\tilde{\tau}_0-\tau)\rp^{2/3},	\label{r-tau-a}		\ee
where $\tilde{\tau}_0$ is a constant, equal to the value of $\tau$ at which the particle reaches the black hole singularity.  Physical quantities associated with these geodesics are used as coordinates in the Lema\^itre system
\cite{Lem33}, beginning with the proper time $\tau$.  However, as discussed in Section \ref{sec:PGtime}, for an infalling $\tilde{E}=1$ observer, $\Delta \tau = \Delta t_{PG}$.  Therefore we can substitute $\tau_0-t_{PG}$ for $\tilde{\tau}_0-\tau$ in (\ref{r-tau-a}), yielding
\be
r[t_{PG}]=\lp\frac{3}{2}\sqrt{2m}(\tau_0-t_{PG})\rp^{2/3}.	\label{r-rho-a}	\ee

The quantity $\tau_0$, the value of PG time at which a given particle reaches $r=0$, can be used to label the geodesic for that particle.  This $\tau_0$ is then promoted to a coordinate $``\rho"$ so that
\bea
r &=& r[t_{PG},\rho]=\lp\frac{3}{2}\sqrt{2m}(\rho-t_{PG})\rp^{2/3},	\label{r-rho-b}	\\
dr &=& 
-\sqrt{2m/r}(dt_{PG}-d\rho).		\label{dr-dt-drho}
\eea
Radially infalling $\tilde{E}=1$ geodesics have constant values of $\rho$, and thus $\rho$ plays the role of a comoving radial coordinate (that can take negative values).  The Lema\^itre coordinates are \{$ t_{PG},\rho,\theta,\phi $\}, and curves of constant $\rho$ (which delineate the infalling geodesics) are presented in Figure \ref{Fig-rho}.  When (\ref{dr-dt-drho}) is substituted into the PG metric (\ref{PGmetric}), the result is \cite{GH78,Blau14}	 
\bea	ds^2 &=& -dt_{PG}^2 +\frac{2m}{r[t_{PG},\rho]}d\rho^2 + r^2[t_{PG},\rho]d\Omega^2,	\label{lemaitremetric} 
\eea
where $r[t_{PG},\rho]$ is given explicitly by (\ref{r-rho-b}). 
\twofigures{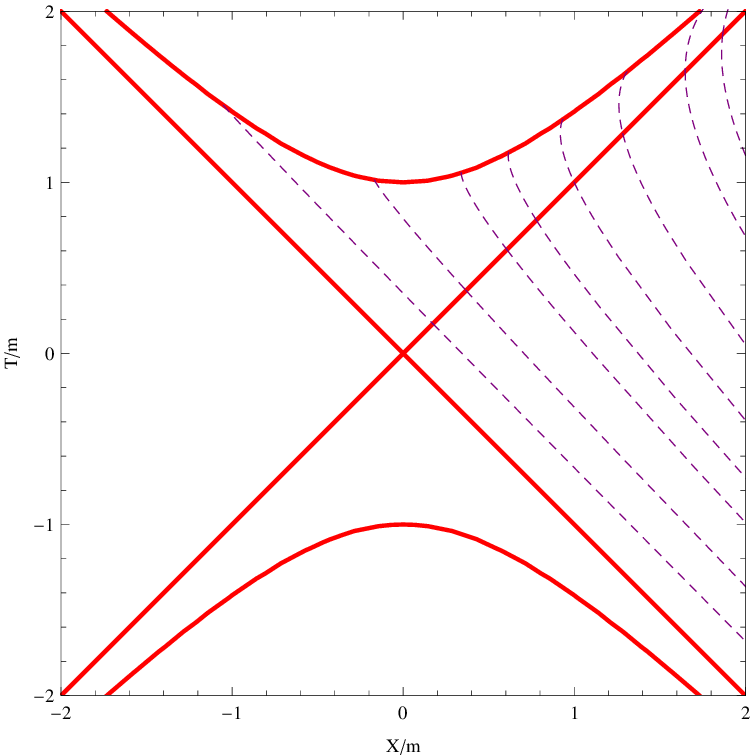}{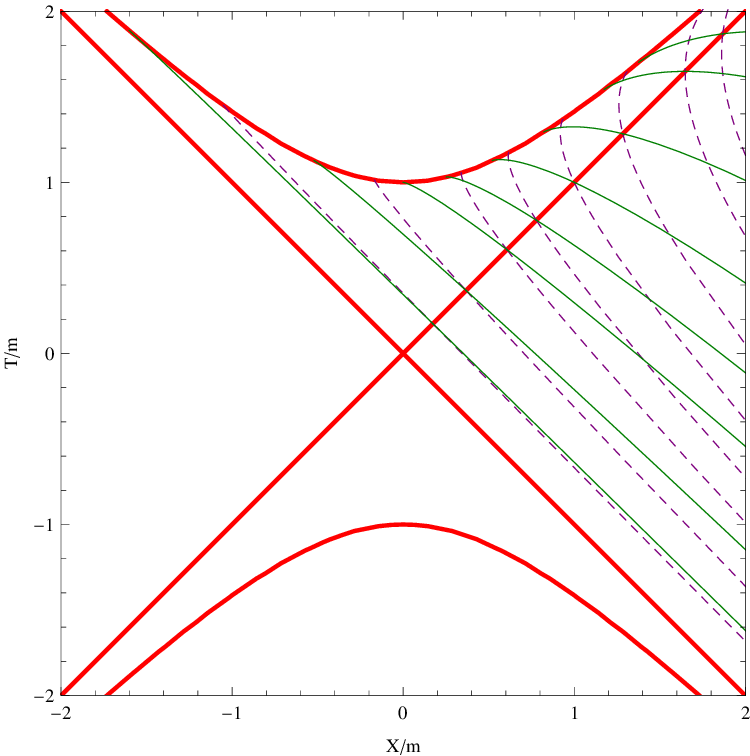}{Top panel: Dashed curves of constant $\rho$, i.e. infalling radial geodesics. From the left, these curves denote the cases $\rho=\{-11m/3,-2m/3,4m/3,7m/3,10m/3,13m/3,16m/3,35m/6\}$.  Bottom panel:  The same geodesics (dashed) from the top panel, now shown together with (thin solid) curves of constant $t_{PG}$ .  The curves of constant PG time, correspond to $t_{PG}=\{-5m,-2m,0,m,2m,3m,4m,4.5m\}$. Each of these $t_{PG}$ curves has a corresponding geodesic that intersects it precisely at $r=2m$; such a geodesic satisfies $\rho = t_{PG}+4m/3$. } 
The coordinate $t_{PG}$ is always timelike, $\rho$ is always spacelike, and there is no coordinate singularity at the horizon.  Thus the Lema\^itre system provides a \emph{diagonal} representation of the Schwarzschild geometry that is well-behaved everywhere outside the physical singularity at $r=0$.  However, this has come at the expense of giving the metric explicit time dependence.   

Now, the volume within the black hole with respect to this slicing is the region along slices of constant $t_{PG}$ between $r[t_{PG},\rho]=0$ and $r[t_{PG},\rho]=2m$.  For this calculation the range of $t_{PG}$ and $\rho$ will be taken as $(-\infty,\infty)$, indicating that the first geodesics reached $r=0$ at very large negative values of $t_{PG}$.  We also have 
\be
\sqrt{ {}^{(3)}g } = \sqrt{2m}\,r^{3/2}[t_{PG},\rho]\sin\theta = 3m(\rho-t_{PG})\sin\theta.		\ee
Hence we obtain
\bea 
\mbox{Vol}_{Lemaitre} &=& \int^{2\pi}_{0} \int^{\pi}_{0}\int^{r[t_{PG},\rho]=2m}_{r[t_{PG},\rho]=0} \sqrt{ {}^{(3)}g } \, d\rho \, d\theta \, d\phi \non \\
&=& 4\pi\int^{r[t_{PG},\rho]=2m}_{r[t_{PG},\rho]=0} 3m(\rho-t_{PG})\, d\rho \, .		\label{LemaitreVol-a}
\eea
It is pleasantly straightforward to see from (\ref{r-rho-b}) that $r[t_{PG},\rho]=0$ corresponds to $\rho=t_{PG}$, and $r[t_{PG},\rho]=2m$ corresponds to $\rho = t_{PG}+4m/3$; the latter relation is illustrated graphically in the bottom panel of Figure \ref{Fig-rho}.  Consequently,
\bea 
\mbox{Vol}_{Lemaitre}  &=&  12\pi m\int^{t_{PG}+4m/3}_{t_{PG}} (\rho-t_{PG}) \, d\rho 	\\
&=& 6\pi m (\rho^2-2\rho \, t_{PG}) \bigg|^{\rho=t_{PG}+4m/3}_{\rho=t_{PG}}	\non
=\frac{32\pi}{3}m^3.  
\eea 
\vskip 0.3cm
The time dependence has been completely eliminated, leaving precisely the volume obtained in (\ref{PGVol}).  Even though this result merely confirms what one would expect based on the results of Section \ref{sec:PGtime}, it is still satisfying to see it arise from a computation with both a time-dependent integrand and time-dependent limits of the integral.
\section{Relating gradients of time functions to four-velocity} 
\label{four-velocity}
Let $u^{a}$ represent the four-velocity of an observer falling inward on a radial geodesic with energy per unit mass $\tilde{E}=1/\sqrt{p}$.  This motion satisfies, for any $p>0$,	
\be \frac{dr}{d\tau} = -\sqrt{\frac{1}{p}-\lp 1- \frac{2m}{r} \rp } \, . 
\ee	
In LMP-family coordinates (i.e. either $t_{LMP}$ or $t_{LMP*}$), we have 
\bea \label{u_LMP-appendix}
u^a &=& \lp \sqrt{p}, -\sqrt{\frac{1}{p} - \lp 1- \frac{2m}{r} \rp } \, , 0, 0  \rp, 	\\ 
u_{a} &=&\lp  -\frac{1}{\sqrt{p}},  0, 0, 0 	\rp \, . \label{u_LMP_lower} 
\eea
In proper time-family coordinates (i.e. either $t_{DF}$ or $t_{HF}$), we have 
\bea \label{u_DF-appendix}
u^a &=& \lp 1, -\sqrt{\frac{1}{p} - \lp 1- \frac{2m}{r} \rp } \, , 0, 0  \rp, 	\\ 
u_{a} &=& \lp  -1,  0, 0, 0 	\rp \, . 	\label{u_DF_lower} 
\eea
In reference \cite{MP01} it was shown for the $p<1$ case that one can relate $u_{a}$ to the gradient of a time function, in this case $t_{LMP}$:  
\be 
u_{a} = -\frac{1}{\sqrt{p}}\partial_{a} t_{LMP}. \label{u--LMP}\ee
This is consistent with (\ref{u_LMP_lower}), but more general, since it holds in any coordinate system.

Since $ t_{HF} = t_{LMP} / \sqrt{p}$ we also have
\be u_{a}= -\partial_{a} t_{HF} \, .	\label{u--HF}
\ee
If one considers instead an observer with $p>1$, the same reasoning leads to 
\be u_{a} =	 -\partial_{a} t_{DF}	 \label{u--DF}\, , \ee 
and since $t_{LMP*}^{(p)} = \sqrt{p}\, {t_{DF}}^{(p)}$ we are led to
\be
u_{a} = -\frac{1}{\sqrt{p}}\partial_{a} t_{LMP*}\ .	\label{u--LMP*}	
\ee		
Comparison of (\ref{u--LMP}) with (\ref{u--LMP*}), and (\ref{u--HF}) with (\ref{u--DF}) illustrates how $t_{LMP*}$ is the $p>1$ analog of $t_{LMP}$, and $t_{DF}$ is the $p>1$ analog of $t_{HF}\,$. 	

\end{document}